\documentclass[final,5p,times,twocolumn,authoryear]{elsarticle}

\usepackage{amssymb}
\usepackage{amsmath}
\usepackage{lipsum}
\usepackage{hyperref} 
\usepackage{orcidlink}

\journal{Physics Letters B}

\begin{document}
\begin{frontmatter}

\title{Impact parameter and transverse charge densities of the pion and kaon
}

\author[first,third]{Fernando Chandra\orcidlink{0009-0003-8583-4054}}
\ead{fci731@uregina.ca}
\affiliation[first]{organization={Department of Physics, Faculty of Graduate Studies and Research (FGSR), University of Regina},
            city={Regina},
            postcode={S4S 0A2}, 
            state={Saskatchewan},
            country={Canada}}
\author[second]{Parada~T.~P.~Hutauruk\orcidlink{0000-0002-4225-7109}}
\ead{phutauruk@hiroshima-u.ac.jp}
\affiliation[second]{organization={International Institute for Sustainability with Knotted Chiral Meta Matter (WPI-SKCM$^2$), Hiroshima University}, 
            addressline={1-3-1 Kagamiyama},
            city={Higashi-Hiroshima},
            postcode={739-8531}, 
            state={Hiroshima},
            country={Japan}}
\author[third]{Terry Mart\orcidlink{0000-0003-4628-2245}}
\ead{terry.mart@sci.ui.ac.id}
\affiliation[third]{organization={Departemen Fisika, Fakultas Matematika dan Ilmu Pengetahuan Alam (FMIPA), Universitas Indonesia},
            city={Depok},
            postcode={16424}, 
            state={Jawa Barat},
            country={Indonesia}}

\begin{abstract}
The impact parameter and transverse charge densities provide spatially resolved information on hadron structure and probe the interplay between dynamical chiral symmetry breaking and SU(3)-flavor symmetry breaking. We investigate the valence-quark distributions of the pion and kaon using their unpolarized vector GPDs at nonzero skewness $\xi$ within the covariant Nambu--Jona-Lasinio model, with ultraviolet divergences regulated by the Schwinger proper-time scheme. Increasing $\xi$ shifts the distributions toward smaller $x$, reduces their central magnitude at $b_\perp=0$, and broadens their transverse profiles. These results provide insight into the spatial structure of the pion and kaon and offer stringent tests of nonperturbative QCD dynamics, with relevance to forthcoming measurements at Jefferson Lab and the Electron-Ion Collider.
\end{abstract}


\end{frontmatter}

\section{Introduction}
\label{introduction}
In quantum chromodynamics (QCD), the pion and kaon are among the lightest hadronic bound states and emerge as pseudo-Goldstone bosons associated with the spontaneous breaking of chiral symmetry~\cite{Zweig:1964ruk,Zweig:1964jf,Lee:1972fj,Hutauruk:2025gnb}. Their properties thus provide important insights into dynamical chiral symmetry breaking (DCSB), quark confinement, and the emergence of asymptotic freedom, as well as into the nonperturbative dynamics of QCD~\cite{Roberts:2021nhw}. Generalized parton distributions (GPDs) offer a comprehensive framework for probing the internal structure of hadrons, unifying information from parton distribution functions (PDFs), generalized form factors (GFFs), and mechanical properties of hadrons, including the pion and kaon~\cite{Diehl:2003ny,Belitsky:2005qn,Guo:2026swz}. These aspects have been investigated extensively using a variety of theoretical and phenomenological approaches, providing insights into meson structure and the manifestations of DCSB and confinement in hadrons~\cite{Roberts:2021nhw,Son:2024uet,Chandra:2025pqs,Chandra:2026smf}.

A particularly useful representation of the spatial structure encoded in the GPDs of the pion and kaon is provided by the impact-parameter space, which describes the transverse spatial distribution of a quark carrying a longitudinal momentum fraction $x$~\cite{Burkardt:2000za,Diehl:2002he}. GPDs therefore provide a powerful framework for connecting theoretical descriptions with experimental observables and for probing the spatial and mechanical structure of hadrons. They depend on three independent kinematic variables: the longitudinal momentum fraction $x$, the skewness parameter $\xi$, and the squared momentum transfer $t$. Through appropriate limits, integrals, and moments in $x$, GPDs encode several fundamental hadronic quantities, including parton distribution functions (PDFs) and generalized form factors (GFFs). In particular, PDFs are recovered in the forward limit $\xi,t\to0$, while the $t$-dependence of GPDs determines the transverse spatial distribution of partons through a Fourier transform to impact-parameter space~\cite{Burkardt:2000za}.

Experimentally, GPDs can be probed through hard exclusive processes, most notably deeply virtual Compton scattering (DVCS) and deeply virtual meson production (DVMP)~\cite{Ji:1996ek,Radyushkin:1997ki,Favart:2015umi}. Their extraction from experimental observables, however, remains challenging because these processes provide access to GPDs only through convolutions with perturbatively calculable hard-scattering kernels. Existing measurements of hard exclusive processes, including DVCS~\cite{HERMES:2001bob,CLAS:2001wjj,ZEUS:2003pwh,H1:2005gdw} and Sullivan-type processes~\cite{Sullivan:1971kd}, have therefore provided important constraints on the spatial and momentum structure of hadrons, while a model-independent determination of GPDs remains difficult. Further experimental progress is expected from next-generation facilities, including the Electron-Ion Collider (EIC)~\cite{Arrington:2021biu}, the Electron-ion Collider in China (EicC)~\cite{Anderle:2021wcy}, the Apparatus for Meson and Baryon Experimental Research (AMBER/COMPASS++) at CERN~\cite{Adams:2018pwt}, the J-PARC extension project and its hadron experimental program~\cite{Sakuma:2022twx,Aoki:2021cqa}, and the proposed 22-GeV upgrade of Jefferson Lab~\cite{Accardi:2023chb,Mart:2026tyg}. These facilities will provide complementary opportunities to constrain GPDs and, in particular, to elucidate the three-dimensional momentum and spatial structure of hadrons.

In this study, we investigate the impact-parameter dependence and transverse charge distributions of the pion and kaon, with particular emphasis on their spatial structure as a function of the longitudinal momentum fraction $x$. The impact-parameter-dependent parton distributions are obtained from the pion and kaon GPDs through a Fourier transform with respect to the transverse momentum transfer. This representation provides information on the transverse spatial distribution of partons relative to the hadron center~\cite{Burkardt:2000za}, linking the longitudinal momentum fraction carried by the active quark and its transverse distance from the meson center. This provides information that ordinary PDFs cannot contain. A particularly important relation is that the transverse width generally decreases as $x \to 1$. Physically, a quark carrying nearly all of the meson's longitudinal momentum tends to be localized closer to the transverse center of momentum. In addition to the impact parameter dependence, we evaluate the transverse charge density, which is obtained from a two-dimensional Fourier transform of the electromagnetic form factor. The charge density gives a direct spatial representation of the electromagnetic structure of the pion and kaon. To do that, we employ the covariant Nambu--Jona-Lasinio (NJL) model, regularized using the Schwinger proper-time scheme to control ultraviolet divergences and incorporating an infrared cutoff to implement quark confinement. The model has been successfully applied to meson form factors (FFs) and parton distribution functions (PDFs)~\cite{Hutauruk:2016sug,Hutauruk:2018zfk,Cloet:2006bq,Hutauruk:2026ywo}. Building on our previous studies~\cite{Chandra:2025pqs,Chandra:2026smf}, we extend the analysis to the transverse spatial structure of the pion and kaon. We calculate the impact-parameter-dependent distributions and transverse charge densities for different values of the skewness parameter $\xi$ and momentum transfer $t$. Particular attention is paid to the contributions of light and strange quarks in the pion and kaon, their dependence on the transverse impact parameter $b_\perp$ and longitudinal momentum fraction $x$, and the effects of SU(3)-flavor-symmetry breaking. This analysis allows us to quantify the differences between the $u$- and $\bar{s}$-quark spatial distributions in the kaon and to contrast them with the corresponding distributions in the pion. This will provide a spatial probe of SU(3) breaking. Furthermore, we also evaluate the impact parameter and charge density of the pion and kaon for nonzero skewness, where the initial and final mesons carry different longitudinal momentum. Consequently, the Fourier transform no longer has a simple probability density interpretation. This is also important because GPDs contain more information than the ordinary PDFs. Furthermore, the results of this study offer a stringent test of nonperturbative QCD dynamics.

\section{Formalism}
\label{sec:formalism}
The impact-parameter-dependent distributions are obtained by Fourier transforming the GPDs with respect to the transverse momentum transfer. Before discussing the impact-parameter representation, we first outline the calculation of the pion and kaon GPDs within the covariant Nambu--Jona-Lasinio (NJL) model. The generalized parton distribution $\mathcal{H}(x,\xi,t)$ depends on three independent kinematic variables: the longitudinal momentum fraction $x$, the skewness parameter $\xi$, and the squared four-momentum transfer $t$. Within the Schwinger proper-time-regularized NJL model, the vector GPDs can be expressed as~\cite{Chandra:2026smf}
\begin{eqnarray}
    \label{eq1}
    \mathcal{H}_M^q (x,\xi,t) 
    &=& \frac{N_C g_{M qq}^2}{8\pi^2} \int_{\tau_{\mathrm{UV}}^2}^{\tau_{\mathrm{IR}}^2} \frac{d\tau}{\tau} \Theta_{01} \exp \big[ -\tau \big( \Delta_1 \big) \big] \nonumber \\ 
    &+& \frac{N_C g_{M qq}^2}{8\pi^2} \int_{\tau_{\mathrm{UV}}^2}^{\tau_{\mathrm{IR}}^2} \frac{d\tau}{\tau} \Theta_{23} \exp\big[ -\tau \big( \Delta_2 \big)  \big] \nonumber \\
    &+& \frac{N_C xg_{{M} qq}^2 }{8\pi^2 \xi} \int_{\tau_{\mathrm{UV}}^2}^{\tau_{\mathrm{IR}}^2} \frac{d\tau}{\tau} \Theta_{45} \exp\big[ -\tau \big( \Delta_0 \big) \big] \nonumber \\
    &+& \frac{N_C g_{{M} qq}^2}{16\pi^2\xi}  \Theta (\alpha_6) \Theta (\alpha_7) \big[(1-x)t +  B \big] \nonumber \\
    &\times& \int_{\tau_{\mathrm{UV}}^2}^{\tau_{\mathrm{IR}}^2} d\tau \int_0^{1}d{\beta}  \,\Theta\left(1-\beta-\beta_1\right) \nonumber \\
    &\times&\exp\big[ -\tau \big( \Delta_3 \big) \big], 
\end{eqnarray}
where $N_c=3$ is the quark color number, $g_{Mqq}$ is the meson-quark coupling, the the subscript $M$ denotes the meson, with $M=K$ for the kaon and $M=\pi$ for the pion. The GPD expression in Eq.~(\ref{eq1}) is written for generic constituent-quark masses and is therefore directly applicable to the kaon. For the pion, the corresponding expression follows by setting the constituent-quark masses equal, $M_s \rightarrow M_u$ (or, equivalently, $M_q \rightarrow M_{q'})$. The remaining quantities are defined as $\Theta_{01} = \Theta (\alpha_{0} ) \Theta ( \alpha_1 ), \Theta_{23} = \Theta (\alpha_2 )  \Theta ( \alpha_3), \Theta_{45} = \Theta (\alpha_4)  \Theta (\alpha_5), \Delta_1 = M_{q'}^2 - \alpha _1 (M_{q'}^2-M_q^2)-\alpha_1 (1-\alpha_1) m_{M}^2, \Delta_2 = M_{q'}^2 - \alpha _2 (M_{q'}^2-M_q^2)-\alpha_2 (1-\alpha_2) m_{{M}}^2, \Delta_0 = M_q^2 - \beta_0 (1-\beta_0) t, \Delta_3 = M_q^2-\beta(M_q^2-M_{q'}^2) - {\beta} (1-{\beta})m_{{M}}^2 - \beta_1 (1-\beta_1 -{\beta}) t$, and $B =  2x\big[m_{{M}}^2- (M_{q}-M_{q'})^2\big]$.
Furthermore, the quantities entering Eq.~(\ref{eq1}) are defined as $\alpha_0 = (x+\xi)/(1+\xi), \alpha_1 = (1-x)/(1+\xi), \alpha_2 = (x-1)/(\xi-1), \alpha_3 = (\xi-x)/(\xi-1), \alpha_4 = 1-(x/\xi),  \alpha_5 = 1+(x/\xi), \alpha_6 = (\xi+x-(1+\xi)\beta)/\xi, \alpha_7 = (\xi-x+(1-\xi)\beta)/\xi, \beta_0 = 0.5\alpha_5, \beta_1 = 0.5\alpha_7$.
We emphasize that the pseudoscalar-meson GPDs in Eq.~(\ref{eq1}) are formulated for arbitrary quark masses. In this work, we specialize to the pion and kaon and evaluate their GPDs at nonzero skewness, with the scale dependence evolved to $\mu^2=4$ and $27~\mathrm{GeV}^2$ using the NLO DGLAP evolution equations~\cite{Miyama:1995bd}.

To determine the impact-parameter space of the kaon and pion, we apply the Fourier transform of the transverse momentum of the GPD by choosing $t = -\Delta_\perp^2$. Hence, the expression can be written as
\begin{eqnarray}
    \label{eq:impact}
    q_M(x,\xi,b_\perp) &=& \int^\Lambda_0 \frac{d^2\Delta_\perp}{(2\pi)^2} e^{-i\Delta_\perp b_\perp} \mathcal{H}_M^q(x,\xi,-\Delta_\perp^2),
\end{eqnarray}
where $\Lambda$ denotes the upper integration limit, chosen to ensure convergence, following the procedure of Ref.~\cite{Burkardt:2000za}, and $b_\perp$ represents the transverse quark distance from the center of meson momentum.

The transverse charge density of the pion and kaon is obtained from the corresponding spacelike electromagnetic FFs through a two-dimensional Fourier transform, which reduces to a radial Fourier--Bessel transform. It is defined as
\begin{eqnarray}
    \rho_M (b_\perp) &=&  \int_0^1dx\,q_M(x,\xi,b_\perp).
\end{eqnarray}
This expression can equivalently be written in terms of the corresponding meson form factor as
\begin{eqnarray}
 \rho_M (b_\perp) &=& \int_0^\infty \frac{dQ}{2\pi} Q j_0(Qb_\perp) F_{M}(Q^2),
\end{eqnarray}
where $Q^2 = -q^2$ is the four-momentum squared of the virtual photon, $j_0(x)$ is the Bessel function, $F_M(Q^2)$ is the meson form factor, and the transverse charge density is normalized according to $\int d^2b_\perp\,\rho_M(b_\perp)=1$, consistent with Refs.~\cite{Burkardt:2000za,Burkardt:2002hr,Miller:2010nz}.

\section{Numerical results and discussion}
\label{sec:results}
The numerical results of the impact-parameter distributions of the pion and kaon are shown in Figs.~\ref{fig1}--\ref{fig3}, while the transverse charge densities of the mesons and their constituents are presented in Figs.~\ref{fig4}--\ref{fig9}. In the present work, the calculations employ the same parameter set as those used in Refs.~\cite{Chandra:2025pqs,Chandra:2026smf}.

\subsection{Pion and kaon impact-parameter-dependent GPDs}
\begin{figure}[t]
	\centering 
    \includegraphics[width=0.239\textwidth, angle=0]{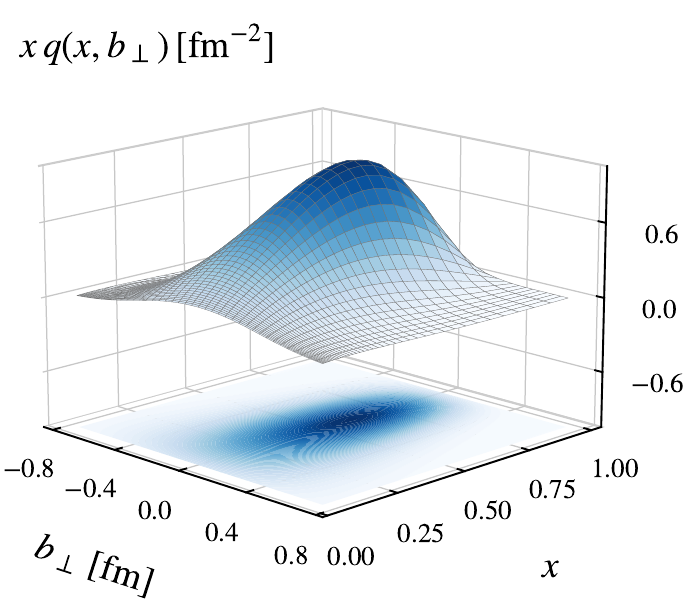}
    \includegraphics[width=0.239\textwidth, angle=0]{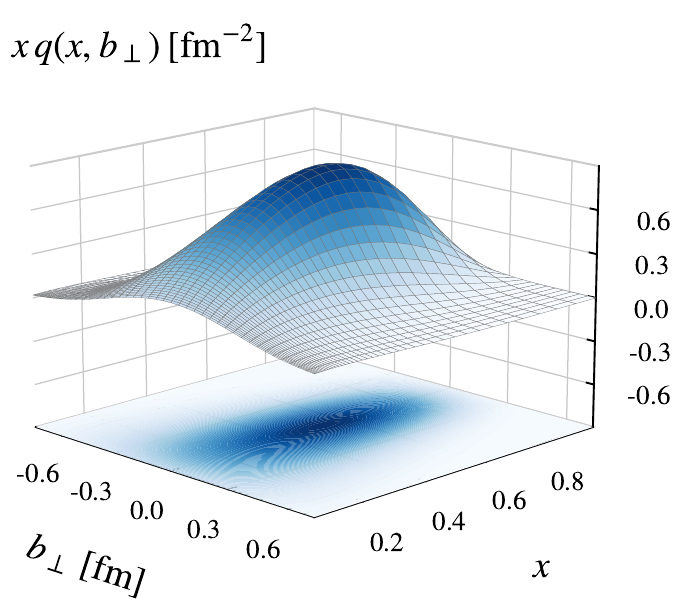}\\
    \includegraphics[width=0.48\textwidth, angle=0]{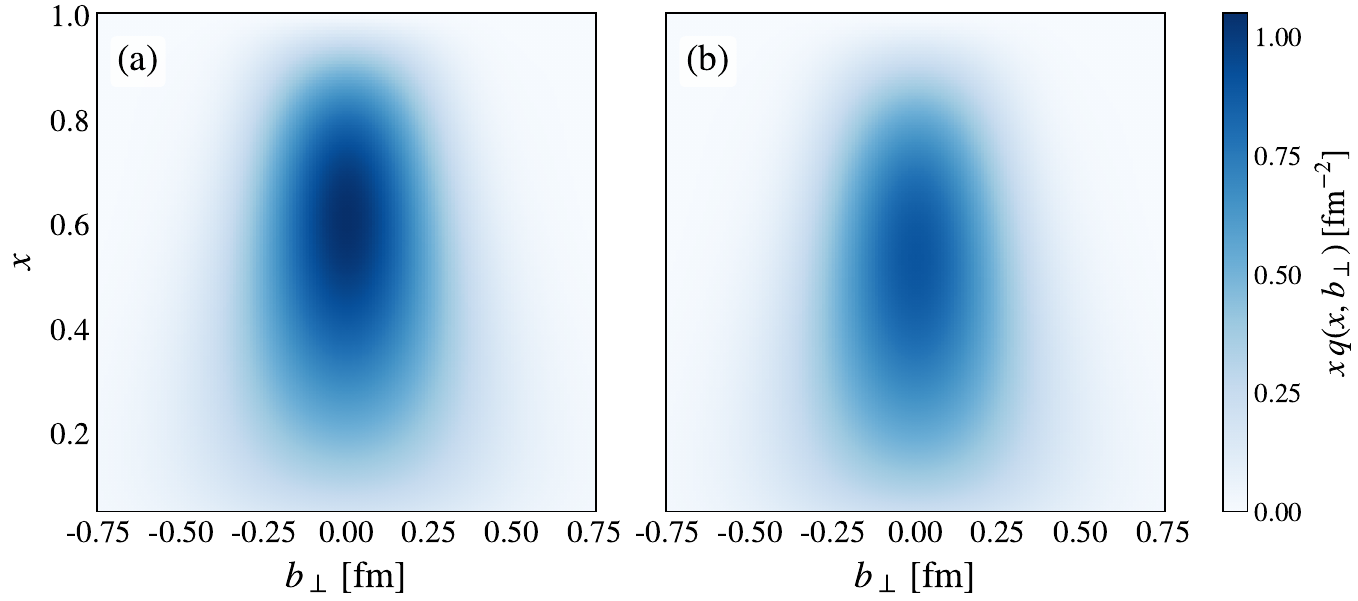}
	\caption{Impact-parameter-dependent GPDs of the pion's valence quark,  multiplied by the longitudinal momentum $x$ as functions of $x$ and $b_\perp$ (Upper panels) and their projections in the $x$ and $b_\perp$ plane (Lower panels)  at scales (a) $\mu^2 =$ 4 GeV$^2$, and (b) $\mu^2 =$ 27 GeV$^2$.} 
	\label{fig1}%
\end{figure}

The three-dimensional pion impact-parameter-dependent distribution multiplied by the longitudinal momentum $x$ and its projection at $\mu^2=4$ and $27~\mathrm{GeV}^2$ are shown in Fig.~\ref{fig1}. As seen in Fig.~\ref{fig1}(a), at $\mu^2=4~\mathrm{GeV}^2$ the distribution exhibits a pronounced peak around $x\simeq0.6$ located at the center of the pion $b_\perp=0$, corresponding to the center of the transverse momentum plane. This behavior is also clearly reflected in the corresponding projection of the impact-parameter distribution. 

The pion impact-parameter-dependent distribution at $\mu^2=27~\mathrm{GeV}^2$ is shown in Fig.~\ref{fig1}(b). Compared with the distribution at $\mu^2=4~\mathrm{GeV}^2$, the peak located at $b_\perp=0$ shifts toward lower $x$, from $x\simeq0.6$ to $x\simeq0.5$. The distribution also exhibits a reduced magnitude at the larger scale, as reflected by the corresponding projection in Fig.~\ref{fig1}(a) and \ref{fig1}(b). This scale dependence indicates that the pion's transverse spatial structure broadens and redistributes under QCD scales. Our results on the pion valence quark impact parameter-dependent GPDs are consistent with those found in Ref.~\cite{Adhikari:2021jrh}.
\begin{figure}[t]
	\centering 
    \includegraphics[width=0.239\textwidth, angle=0]{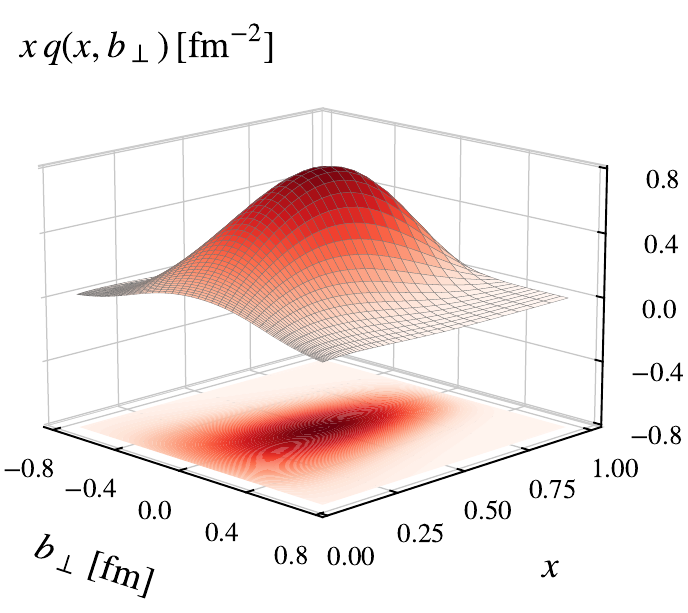}
    \includegraphics[width=0.239\textwidth, angle=0]{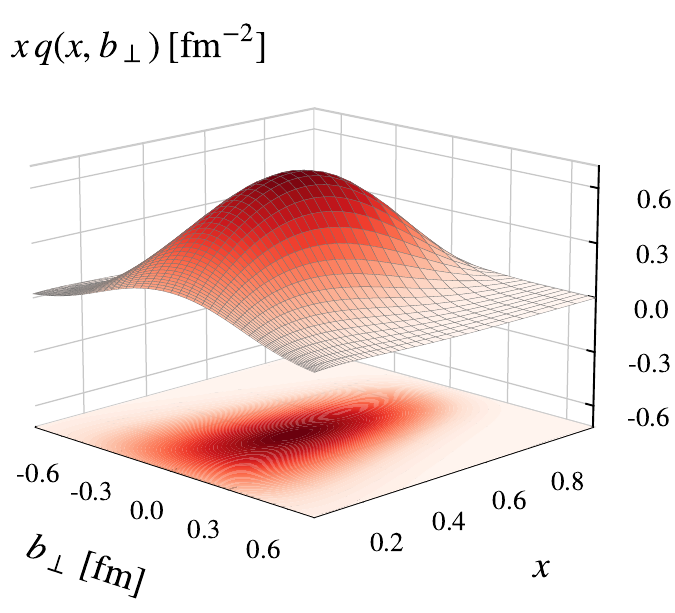}\\
    \includegraphics[width=0.48\textwidth, angle=0]{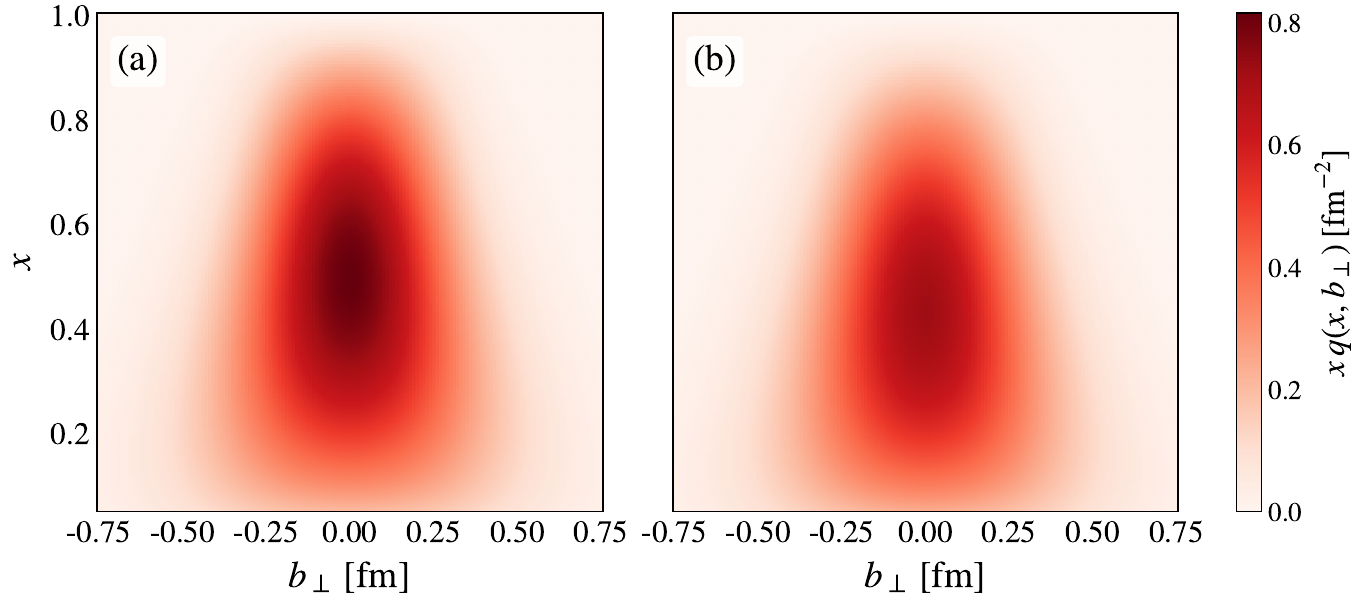}
	\caption{Impact-parameter-dependent GPDs of the kaon's up valence quark, multiplied by $x$ as functions of $x$ and $b_\perp$ (Upper panels) and their projections (Lower panels) at scales (a) $\mu^2 =$ 4 GeV$^2$, and (b) $\mu^2 =$ 27 GeV$^2$.} 
	\label{fig2}%
\end{figure}

In addition to the pion results, we calculate the three-dimensional kaon impact-parameter-dependent GPDs as functions of $x$ and $b_\perp$, together with their projections, at $\mu^2=4$ and $27~\mathrm{GeV}^2$, as shown in Fig.~\ref{fig2}(a) and (b). At $\mu^2=4~\mathrm{GeV}^2$, the upper panel of Fig.~\ref{fig2}(a) shows that the kaon $u$-valence-quark impact-parameter distribution exhibits a peak at $x\simeq0.45$ and $b_\perp=0$. This peak is also clearly visible in the corresponding projection in the lower panel. At the same scale, the peak occurs at a lower $x$ value than that of the pion valence-quark distribution. 

The kaon up-valence-quark impact-parameter distribution at $\mu^2=27~\mathrm{GeV}^2$ is shown in Fig.~\ref{fig2}(b). Compared with the result at $\mu^2=4~\mathrm{GeV}^2$, the peak is slightly reduced in magnitude and shifts toward lower $x$, occurring at $x\simeq0.4$ for $b_\perp=0$. This behavior is clearly noticeable in the corresponding projection shown in the lower panel of Fig.~\ref{fig2}(b).
 
Overall, the maxima of the pion and kaon impact-parameter distributions at $b_\perp=0$ shift toward smaller $x$ with increasing evolution scale, reflecting the redistribution of longitudinal momentum induced by QCD evolution.
\begin{figure}[t]
	\centering 
    \includegraphics[width=0.239\textwidth, angle=0]{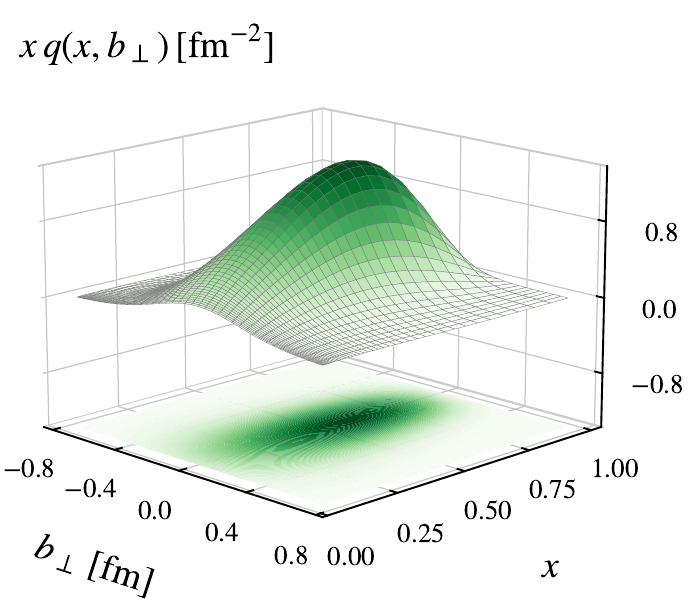}
    \includegraphics[width=0.239\textwidth, angle=0]{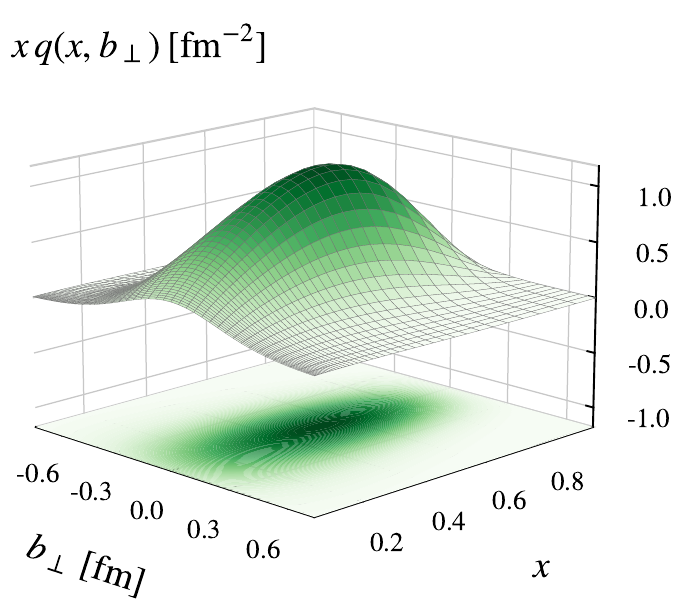}\\
    \includegraphics[width=0.48\textwidth, angle=0]{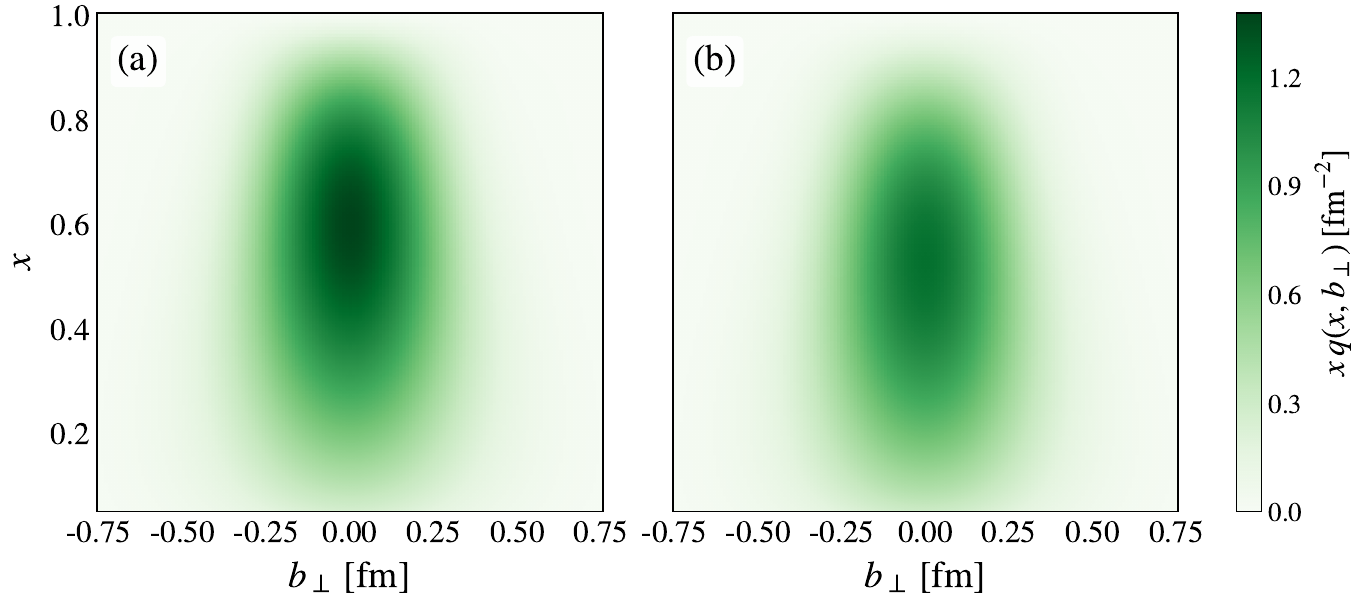}
	\caption{Impact-parameter-dependent GPDs of the kaon's strange valence quark, multiplied by $x$ as functions of $x$ and $b_\perp$ (Upper panels) and their projections (Lower panels) at scales (a) $\mu^2 =$ 4 GeV$^2$, and (b) $\mu^2 =$ 27 GeV$^2$.} 
	\label{fig3}
\end{figure}

We further evaluate the strange-valence-quark impact-parameter distribution of the kaon at $\mu^2=4$ and $27~\mathrm{GeV}^2$, as shown in Fig.~\ref{fig3}(a) and (b), respectively. At $\mu^2=4~\mathrm{GeV}^2$, the distribution exhibits a maximum at $x\simeq0.6$ for $b_\perp=0$, as clearly seen in the projection onto the $x$-$b_\perp$ plane in the lower panel of Fig.~\ref{fig3}(a).

At $\mu^2=27~\mathrm{GeV}^2$, the maximum shifts toward smaller $x$, reaching $x\simeq0.5$ at $b_\perp=0$, while its magnitude decreases with increasing evolution scale, as shown in the lower panel of Fig.~\ref{fig3}(b).

\subsection{Pion and kaon transverse charge densities}
\begin{figure*}[!]
	\centering 
    \includegraphics[width=0.3\textwidth, angle=0]{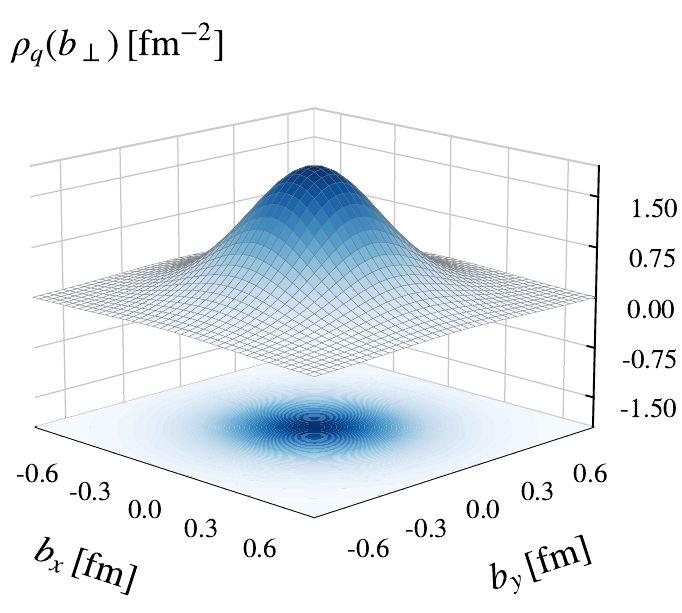}
	\includegraphics[width=0.3\textwidth, angle=0]{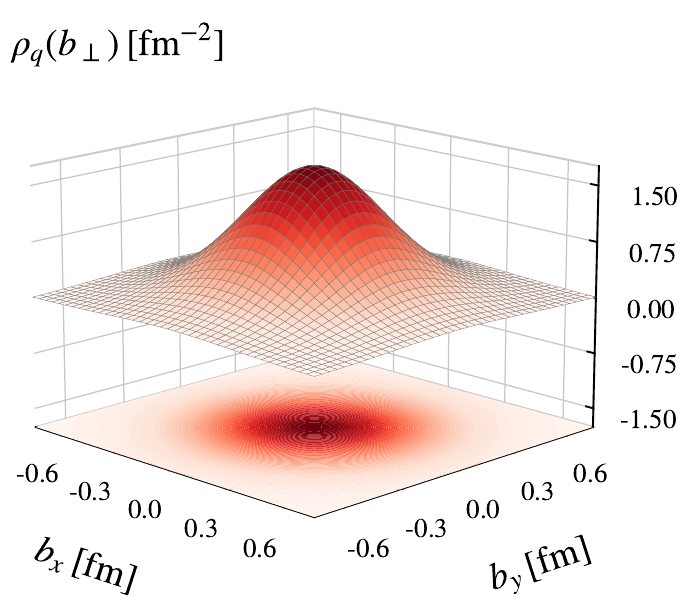}	
    \includegraphics[width=0.3\textwidth, angle=0]{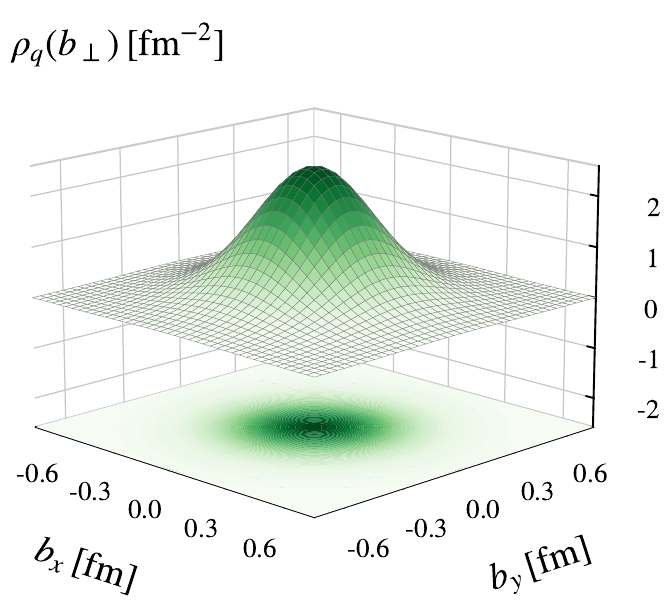} \\
    \includegraphics[width=1.0\textwidth, angle=0]{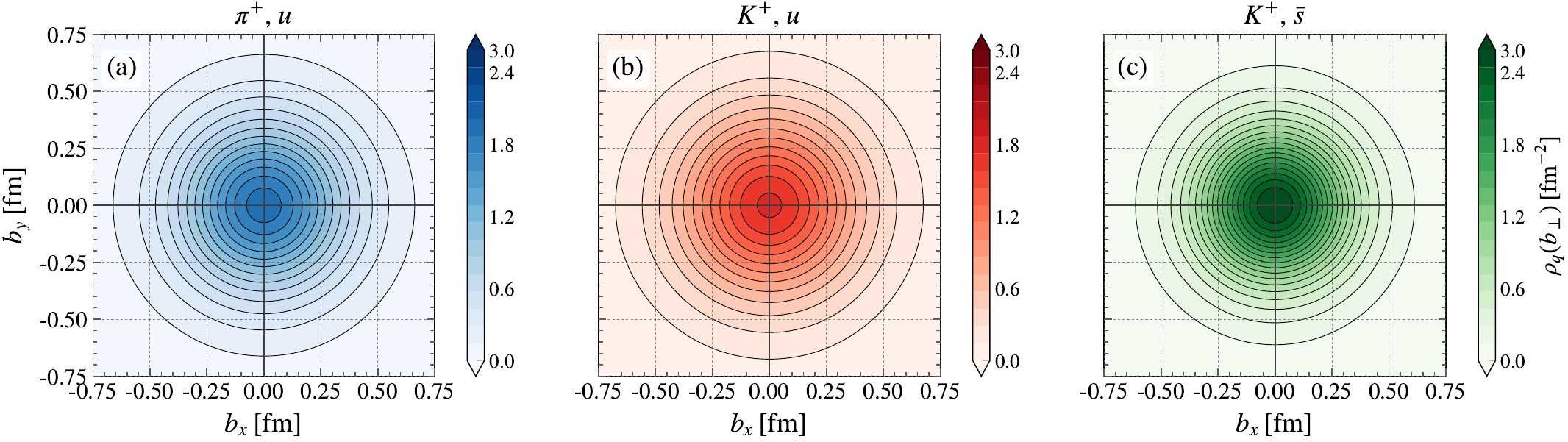}
	\caption{The unpolarized transverse charge probability densities as functions of $b_x$ and $b_y$ (Upper panels) and their projections (Lower panels) at scale $\mu^2 =$ 4 GeV$^2$ for (a) the pion up valence quark (blue), (b) the kaon up valence quark (red), and (c) the kaon strange valence quark (green).} 
	\label{fig4}
\end{figure*}

In Fig.~\ref{fig4}, we show our results for the unpolarized transverse probability density of the pion and kaon as functions of $b_x$ and $b_y$ at $\mu^2 =$ 4 GeV$^2$. Figure~\ref{fig4}(a) shows the results of the transverse probability of the up valence quark of the pion in the $b_x$-$b_y$ projection plane. We find that the unpolarized transverse probability density is symmetric at $b_x =0$ and $b_y =0$, which is exactly located at the center of the pion distribution. This result is compatible with that obtained in Ref.~\cite{Adhikari:2021jrh}, calculated using the Nambu-Jona-Lasinio basis light-front quantization (NJL-BLFQ) model. Such symmetric unpolarized transverse probability density of the pion is also clearly indicated in the lower panel of Fig.~\ref{fig4}(a).

We then evaluate the unpolarized transverse probability density for the up valence quark of the kaon at $\mu^2 = 4$ GeV$^2$, as depicted in Fig.~\ref{fig4}(b). Similar to the pion case, we find that the unpolarized transverse density for the up valence quark of the kaon is symmetrically located at the center of the kaon transverse distribution at $b_x =0$ and $b_y =0$. In comparison to the pion case in Fig.~\ref{fig4}(a), we find that the magnitude of the unpolarized transverse density of the kaon is rather similar to that for the pion. However, the transverse probability density of the kaon has a wider distribution than that for the pion, as indicated in the lower panel of Figs.~\ref{fig4}(a) and (b). Again, our finding on the unpolarized transverse probability density of the kaon valence quark is consistent with the results obtained in Ref.~\cite{Adhikari:2021jrh}.

Figure~\ref{fig4}(c) shows the results of the unpolarized transverse probability density for the strange valence quark of the kaon at $\mu^2 = 4$ GeV$^2$. We find that the transverse charge density for the strange quark of the kaon is located at $b_x =0$ and $b_y = 0$, which is also symmetric, like the up quark in the pion and kaon. However, the magnitude of the unpolarized transverse probability density for the kaon strange valence quark is higher than that for the up valence quark of the pion and kaon. Also, the distribution of the kaon strange quark is more concentrated in the center of the $b_x$ and $b_y$ plane, as shown in the lower panel of Fig.~\ref{fig4}(c).
\begin{figure}[t]
	\centering 
    \includegraphics[width=0.45\textwidth, angle=0]{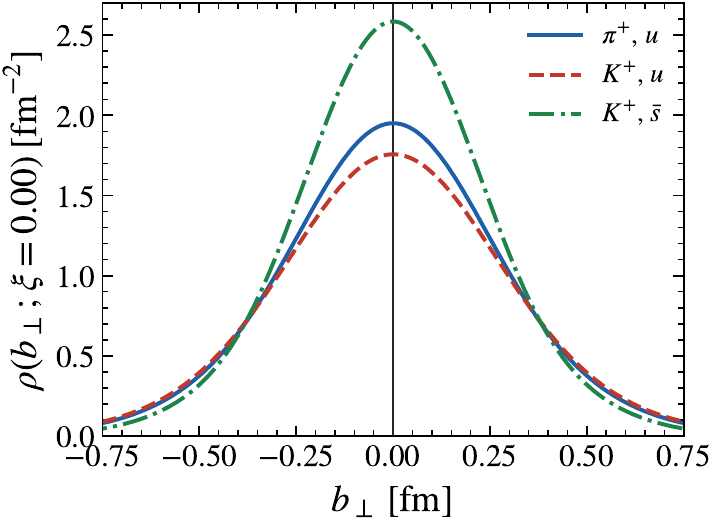}
	\caption{The unpolarized probability densities for the up valence quark of the pion, the up valence quark of the kaon, and the antistrange valence quark of the kaon, calculated with skewness $\xi =0$ at $\mu^2 =$ 4 GeV$^2$ as functions of $b_\perp$.} 
	\label{fig5}%
\end{figure}

Figure~\ref{fig5} shows the unpolarized probability densities of the $u$-valence quark in the pion and kaon and the $\bar{s}$-valence quark in the kaon as functions of the transverse impact parameter $b_\perp$, for $\xi=0$ and $\mu^2=4~\mathrm{GeV}^2$. The $\bar{s}$-quark distribution in the $K^+$ exhibits a substantially larger peak at $b_\perp=0$ than the corresponding $u$-quark distributions in the pion and kaon. It is also more localized in the transverse plane, with a narrower profile, reflecting the larger constituent mass of the strange quark relative to the light quarks. In contrast, the $u$-quark distributions of the pion and kaon have comparable magnitudes and transverse widths over most of the $b_\perp$ range, with noticeable differences emerging only near the center. This behavior is consistent with the findings of Ref.~\cite{Adhikari:2021jrh} and provides a clear manifestation of SU(3)-flavor-symmetry breaking in the transverse spatial distributions.
\begin{figure}[t]
	\centering 
    \includegraphics[width=0.45\textwidth, angle=0]{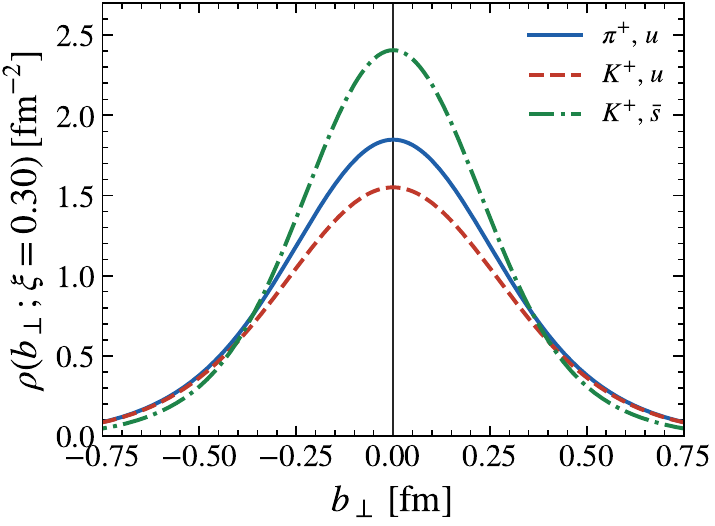}
	\caption{Same as in Fig.~\ref{fig5}, but for the skewness $\xi = 0.3$.} 
	\label{fig6}
\end{figure}

Figure~\ref{fig6} presents the unpolarized probability densities for the $u$-valence quark in the pion and kaon, together with the $s$-valence-quark distribution in the kaon, at $\xi=0.3$ and $\mu^2=4~\mathrm{GeV}^2$. Relative to the $\xi=0$ results in Fig.~\ref{fig5}, the distributions are suppressed near $b_\perp=0$ at nonzero skewness $\xi = 0.3$. Their transverse widths, however, remain comparable across the three cases.
\begin{figure}[t]
	\centering 
    \includegraphics[width=0.45\textwidth, angle=0]{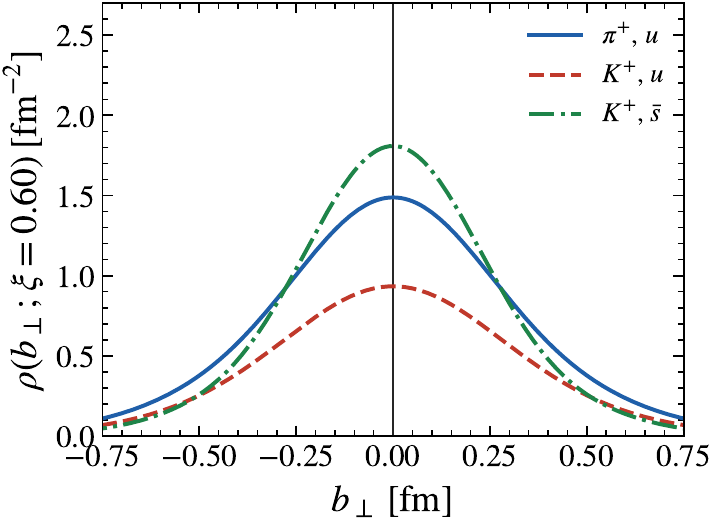}
	\caption{Same as in Fig.~\ref{fig5}, but for the skewness $\xi = 0.6$.} 
	\label{fig7}%
\end{figure}

In Fig.~\ref{fig7}, we show the unpolarized density for the up quarks of the pion and kaon and the strange quark of the kaon at $\mu^2 =$ 4 GeV$^2$ for $\xi = 0.6$. We find that the unpolarized probability density for the up quark of the pion and kaon and the strange quark of the kaon for higher $\xi$ is more suppressed, particularly at the center $b_\perp = 0$, relative to that for the skewness $\xi =0$. Also, we find that the width of the distribution for the pion and kaon up valence quarks and the kaon strange valence quark for $\xi =0.6$ is broader than that for $\xi =0$ and $\xi =0.3$. The broader transverse distribution of $\rho_M(b_\perp,\xi=0.3)$ is expected to arise from the asymmetric initial and final meson momenta at nonzero skewness.
\begin{figure}[t]
	\centering 
    \includegraphics[width=0.45\textwidth, angle=0]{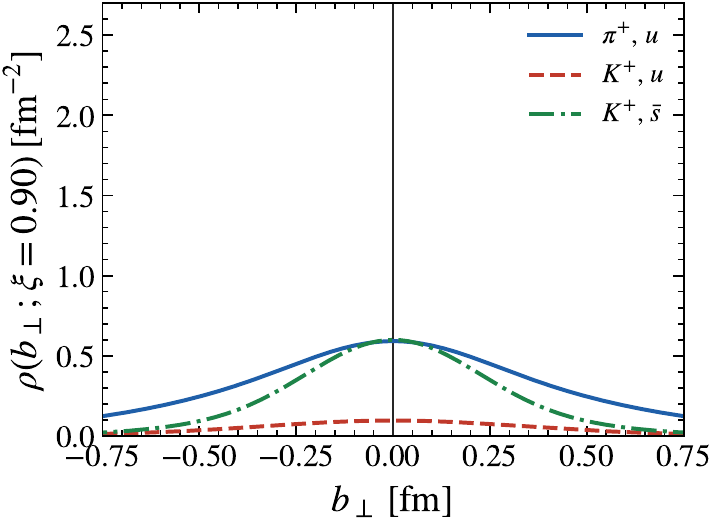}
	\caption{Same as in Fig.~\ref{fig5}, but for the skewness $\xi = 0.9$.} 
	\label{fig8}%
\end{figure}

We further examine the unpolarized transverse densities of the $u$-quark in the pion and kaon and the $\bar{s}$-quark in the kaon at $\mu^2=4~\mathrm{GeV}^2$ and $\xi=0.9$, as shown in Fig.~\ref{fig8}. At this larger skewness, the distributions have substantially smaller peak magnitudes and broader transverse profiles than those obtained at $\xi=0.0, 0.3$, and $0.6$. This behavior reflects the increasing asymmetry between the initial- and final-state meson momenta as $\xi$ increases. 
\begin{figure}[t]
	\centering 
    \includegraphics[width=0.45\textwidth, angle=0]{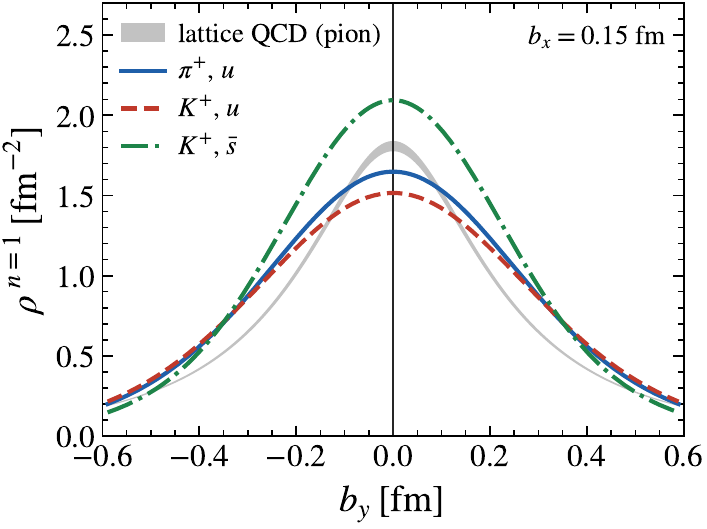}
	\caption{Unpolarized probability densities at fixed $b_x = 0.15$~fm as functions of $b_y$, for the up valence quark in the pion, the up valence quark in the kaon, and the antistrange valence quark in the kaon, evaluated at skewness $\xi = 0$ and scale $\mu^2 = 4$~GeV$^2$. Result for the pion is compared with the lattice QCD simulation of Ref.~\cite{Brommel:2008}. Here $\rho^{n=1}$ denotes the unpolarized transverse density obtained from the first Mellin moment of the $\pi^+$ GPDs.} 
	\label{fig9}%
\end{figure}

Finally, we compare our results for the unpolarized transverse density of the up valence quark in the pion and kaon, and the strange valence quark in the kaon, at $\mu^2 = 4~\mathrm{GeV}^2$ and fixed $b_x = 0.15$~fm, with the lattice QCD simulation result for the pion, as illustrated in Fig.~\ref{fig9}. We find that our result for the pion unpolarized transverse density is comparable in distribution shape with the lattice QCD simulation result~\cite{Brommel:2008}. However, we find different behavior at the peak of the pion transverse density at the center $b_y =0$ and in the transverse distribution at around $b_y \simeq (-0.4$-$-0.2$) fm and at around $b_y \simeq (0.2$-$0.4$) fm. It is worth noting that the lattice QCD result~\cite{Brommel:2008} is simulated using the two flavors of nonperturbatively improved Wilson fermions, with the pion masses $m_\pi =$ 400 MeV in volumes up to $L^3 =(2.1~ \mathrm{fm})^3$ and lattice spacings below 0.1 fm.

\section{Summary and conclusions}
\label{sec:summary}
In summary, we have investigated the unpolarized valence-quark GPDs of the pion and kaon in impact-parameter space, together with their corresponding transverse probability densities, within the covariant Nambu--Jona-Lasinio model, employing Schwinger proper-time regularization to implement quark confinement. These distributions were evaluated at $\mu^2 = 4~\mathrm{GeV}^2$ for several values of the skewness parameter $\xi$, enabling comparison with the available lattice-QCD simulation results~\cite{Brommel:2008}. Such comparisons are particularly valuable in the absence of direct experimental measurements of these quantities.

We found that the maximum peak of the pion and kaon impact parameter distribution at $b_\perp =0$ shifts toward smaller $x$ as the renormalization scale increases. This reflects the redistribution of the longitudinal momentum induced by the QCD evolution.

Furthermore, we found that the peak of the unpolarized probability densities for the $u$-valence quark in the pion and kaon, together with the $s$-valence quark transverse distribution in the kaon at $\mu^2 =$ 4 GeV$^2$ around the center $b_\perp = 0$, decreases, and the width of the transverse distribution becomes broader with increasing the skewness parameter $\xi$. This behavior is expected to arise from the asymmetric initial and final meson momenta at nonzero skewness.

Our result for the pion unpolarized probability density is comparable to the lattice QCD simulation of Ref.~\cite{Brommel:2008}. However, the differences are found in the peak magnitude and transverse distribution around $b_y = (-0.4$-$-0.2$) fm and $b_y = (0.2$-$0.4$) fm.

As demonstrated in this work, the pion and kaon GPDs in the transverse impact-parameter space provide valuable information on the internal structure of the pion and kaon. Although their spin and spatial structure is not directly accessible experimentally, lattice-QCD simulations can provide important constraints on these quantities. GPDs associated with different operators and higher Mellin moments can further resolve the momentum and spatial distributions of quarks within the pion and kaon. A systematic study of these quantities, particularly through lattice simulations, would therefore be valuable for elucidating the detailed internal structure of pions and kaons.

\section*{Acknowledgements}
This work was supported by the World Premier International Research Center Initiative (WPI-SKCM$^2$) of Hiroshima University, MEXT, Japan (P.T.P.H.), and by the PUTI Q1 Grant from University of Indonesia under contract PKS-206/UN2.RST/HKP.05.00/2025 (F.C. and T.M.).

\bibliographystyle{elsarticle-harv}
\bibliography{main}
\end{document}